\documentclass[12pt]{article}

\title{{\bf Lepton pair production in heavy ion collisions and hadronic
phenomenology} }
\author{{Charles Gale}\thanks{Electronic address:
gale@hep.physics.mcgill.ca}
 \vspace*{0.1in} \\
 {\it Physics Department, McGill University}\\ \vspace*{0.2in}
 {\it Montr\'eal, Qu\'ebec, Canada H3A 2T8}\\ \vspace*{0.1in}}

\date{}

\usepackage[dvips]{graphicx}
\begin{document}

\maketitle

\begin{abstract}
We first review the calculations of low invariant mass dilepton
production in relativistic heavy ion collisions, using effective
hadronic Lagrangians. We go through some of
the theoretical techniques used in this kinematical region and
we consider some of the appropriate 
experimental measurements.  Moving up to the intermediate invariant mass
region, we point out some of the uncertainties that show up in theoretical
estimates. Those originate mainly from off-shell effects.  
Finally, as an application of hadronic chiral Lagrangians at finite
temperature, 
we compute the mass shifts and mixing of the $\omega$ and $\phi$
mesons due to scattering from thermal pions.
\end{abstract}

\section*{Introduction}

The field of heavy ion collisions is a very active one, straddling high
energy and nuclear physics. At the upper energy limit of this
flourishing area of research,  the goal is to eventually produce and
study a new
state of matter in the laboratory: the quark-gluon plasma (QGP). That
strongly
interacting matter in conditions of extreme energy densities undergoes a
phase
transition is in fact  a prediction of QCD \cite{qcd}.
To confirm whether the phase transition indeed occurs in relativistic
heavy ion collisions and that the QGP is
formed, one needs a clear signal as a signature for the QGP.
Many approaches have been suggested to elucidate the existence of this
elusive state of matter, but unfortunately, no single measurement can be
singled-out as a ``smoking gun'' candidate. Instead, it appears that
many complementary
experimental data will require simultaneous analysis \cite{harris96}.
One class of observables that appears especially attractive is that of
electromagnetic signals. This owes first to the fact that such probes
essentially suffer little or no final state interactions and thus
constitute reliable carriers able to report on the local conditions at
their emission site.  The calculated emission rates for photons
and lepton
pairs have been shown to strongly depend on the local density and
temperature. Those facts were established some time ago and several
heavy ion experiments specializing in the measurement of 
electromagnetic radiation
are either running right now or being planned. 

In this talk, we specialize on the topic of dilepton measurements
and calculations.  With respect to photons, the extra degree of freedom
associated with a variable invariant mass constitutes an advantage
that is clear when one considers annihilation reactions, for example.
With this in mind, a pioneering experiment was that of the DLS \cite{dls}, 
at the disassembled Bevalac at LBNL. Dileptons were also measured 
by the CERES \cite{ceres} and HELIOS \cite{helios} collaborations
at the SPS at CERN, and will be measured in the PHENIX experiment \cite{cdr}
at RHIC at BNL, and in the HADES experiment at GSI \cite{hades}. 

The emission of lepton pairs from high temperature matter formed
in high energy nucleus$-$nucleus collisions is of great theoretical
and experimental interest.  It can signal a change in the properties of
hadrons (more precisely, correlation functions) in hot hadronic matter
as it approaches a chiral symmetry or a quark deconfinement phase
transition or rapid crossover.  
Vector mesons are prominent in these studies because of their
coupling to the electromagnetic current.  Most studies have focussed
on the lighter ones, $\rho$, $\omega$, and $\phi$, and on the heavier
J/$\psi$.  The best signals are those that have a characteristic shape
or structure.  The $\rho$ meson is very broad in vacuum and undoubtedly
gets even broader at finite temperature.  The J/$\psi$ is narrow; it
has a whole literature of its own within the field.  The $\omega$ and
$\phi$ mesons are rather narrow, but not so narrow that they will
all decay after the hot matter has blown apart in a high energy collision.
This makes them good candidates to study.

\section*{The low mass region}

In general, one can show that the rate for emitting a dilepton pair (or
a real photon, for that matter) is related to the retarded 
photon self-energy, at
finite temperature \cite{photon,gaka91}:
\begin{eqnarray}
E_+ E_- \frac{d R}{d^3 p_+ d^3 p_-}\ = \ \frac{2 e^2}{(2 \pi)^6}\,
\frac{1}{(k^2)^2}  \left[ p_+^{\mu} p_-^{\nu} + p_+^{\nu} p_-^{\mu} -
g^{\mu \nu} p_+ \cdot p_- \right]  Im\, \Pi_{\mu \nu}^R ( k )
\frac{1}{e^{\beta \omega} - 1}\ ,
\end{eqnarray}
where the photon energy is $\omega$, its three-momentum is ${\bf k}$,
and $T = 1/\beta $.

A plausible philosophy then consists of performing an order-by-order  
loop expansion of the 
finite-temperature photon self-energy in the physical fields relevant for 
the
problem at hand. One can then establish a correspondence between this
approach and that consisting of simply using kinetic theory. It
can be shown that the difference in the dilepton rate between those two
scenarios is indeed small up to reasonably high temperatures 
\cite{gaka91}. Systematic evaluation of this sort were carried
out for the theoretical evaluation of real and virtual photon 
emission rates from a
hot gas of mesons \cite{kls,gali94}. We first concentrate on 
the rates for low mass dileptons and recall the assumptions inherent 
to most ``first generation'' estimates. We do not attempt to do justice
to the many calculations in this area: we apologize in advance. In the
following, we highlight some general aspects.

Having in mind the energy regime germane to the CERN SPS, a reasonable
working hypothesis is to first assume that the hadronic 
medium is meson-dominated. 
This statement is experimentally supported \cite{sta96}. 
Also, if thermal equilibrium is indeed established this would influence
the soft sector. This defines a
theoretical framework where finite-temperature hadronic field theory can
be used to evaluate the photon self-energy. Its imaginary part can then
be related to a dilepton emission rate. In practical reality, however,
the above-mentioned connection with kinetic theory is in fact what is
often 
used. Our ``first generation'' dilepton emission rate then contains the
reactions and decays constructed from an ensemble of the lightest
pseudoscalar $(P)$ and vector mesons $(V)$.  Those are $(P)$: $\pi, \eta,
\eta'$, $K$, and $(V)$: $\rho, \omega, \phi$, and $K^*$. The processes
included in our rate calculation are then: $V (P) \rightarrow P (V)
\gamma^*$, $P + P \rightarrow \gamma^*$, $P + V \rightarrow \gamma^*$, and
$V + V \rightarrow \gamma^*$. The respective sizes of those different
contribution to the lepton spectrum are compared and discussed in Ref.
\cite{gali94} (see also Ref.~\cite{ssg96}). 
Note that only two-body reactions were first considered, because of
phase-space arguments. We shall return to this point later. Effective
chiral Lagrangians were used in this calculation,  coupled with 
Vector Meson Dominance \cite{vmd} whenever experimental time-like form 
factors were unavailable. 

A dynamical model is then needed to translate the rates into actual
yields, effectively performing a space-time integration. The rates
obtained by the procedure described above were integrated in a Bjorken
model \cite{ssg96}, and in an hydrodynamical model \cite{soll97}.
Running the results thus obtained through the detector acceptances, one can
compare with the measured data. This has been done by several groups
now, and the results  are nicely summarized in Ref. \cite{drees96}.
Concentrating on the CERES invariant mass coverage, most 
calculations are seen to underestimate the data in the low mass 
sector ($M \approx$ 300 MeV). One
must nevertheless point out the remarkable agreement between the different
theoretical estimates. The latter were obtained within different
dynamical frameworks (see \cite{drees96} and references therein for 
details) and this convergence of theoretical results has been claimed to
show the insensitivity of the CERES results to dynamical modeling. 
Going back to our rates and recalling that
those 
included many reaction channels (even though a few actually dominate), 
it appears that an 
independent verification of those estimates is desirable. In this
context, we focus on two calculations. First,  a calculation
of photon and dilepton production rates from a gas of hot mesons was performed
using chiral reduction formul{\ae} and a virial expansion
\cite{steele96}. The results can be expressed in terms of vacuum
current-current correlation functions, which can be experimentally
constrained. Second, using spectral functions extracted from $e^+ e^-$
data one can thus calculate finite-temperature dilepton emission rates
\cite{huang95}. A comparison between the meson reactions/decays approach
outlined earlier and those two calculations appears in Fig. 1. 
\begin{figure}
\begin{center}
\includegraphics[angle=90, width=10cm]{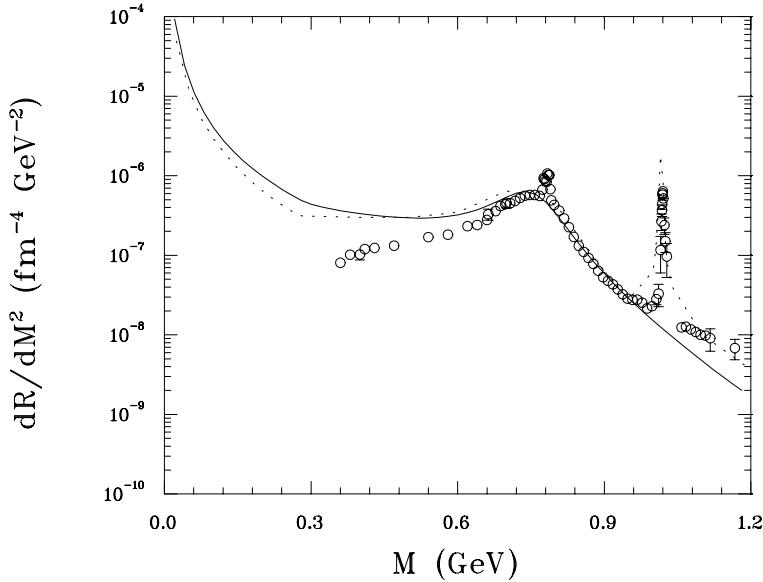}
\end{center}
\caption{\small  A comparison of the dilepton rates at $T$ = 150 MeV  
calculated using
experimentally-extracted spectral functions \cite{huang95} (open
circles), chiral reduction formul{\ae} \cite{steele96} (solid line), and
meson reactions and decays \cite{gali94} (dashed line).}  
\end{figure}
It is important to point out that Ref.~\cite{huang95} will not have the
equivalent of the radiative decay channels, as the analysis is done
starting from $e^+ e^-$ data. Strictly speaking, it will then be relevant
just below the $\rho$ pole, where the pion channels become large. Also,
Ref.~\cite{steele96} does not have the $\phi$ in its formalism: this
explains the absence of the peak. This being said, those three very
different approaches exhibit striking agreement. Bear in mind that {\em
no} arbitrary normalization enters {\em any} of those calculations. This
is not to say that improvements can't be introduced but the unity of
the displayed results is nevertheless satisfying. 

Returning to the underestimation of the low mass data by the majority 
of the theoretical efforts,  several alternative explanations have so far
emerged. Since it is not the point of this talk to address this specific
issue with the detail it deserves, we simply enumerate some of 
them. The early suggestion that the
CERES data could be accounted for by a dropping of the in-medium $\rho$
mass, has aroused immense theoretical interest. This conjecture can 
be incorporated in a model that fits the data \cite{lkbs96}. Another
line of reasoning consists of trying to make more complete dilepton rate
assessments. This, combined with a kinetic theory constrained
by the available data on spectral and multiplicity distributions, 
appears to be a step towards  a resolution of the
discrepancy \cite{mbh96}. Note that those rates will have to be
reconciled with the alternate approaches described on Fig. 1. Finally, 
in-medium baryon effects on the 
$\rho$ spectral function were shown to produce promising results, in
connection with the CERES measurements \cite{rcw97}. The last two 
schools of thought
do not invoke dropping $\rho$ masses. It is obviously too early to pin
down the theoretical explanation which will stand the test of time. However,
the community is eagerly awaiting the dilepton transverse momentum 
spectra which should help to discriminate between theories. 

\section*{The intermediate mass region}

Along with the tantalizing low-mass dilepton excess reported by CERES,
an excess in the intermediate mass region ($m_\phi < M < m_{j/\psi}$) has
also been seen \cite{drees96}. Before 
attempting to make phenomenological estimates, it is best to
exercise caution and to tackle this issue purely theoretically. It has 
been shown in the past that, right past its
threshold, the reaction $\pi + a_1 \rightarrow e^+ e^-$ constituted a
sizeable fraction of the lepton pair spectrum \cite{skg94}. This
exercise was carried out using an effective chiral Lagrangian. However,
the procedure to derive those is unfortunately not unique \cite{meis88}.
Now, because we are no longer in the soft sector, our concern is that
the non-uniqueness of the available chiral Lagrangians will manifest
itself, principally through off-shell effects. With this in mind, we
report here on our findings only briefly, leaving a more 
detailed discussion to be published \cite{gaog97}.

Here, we therefore carry out a mostly pedagogical exercise: what are the
sizes of the possible off-shell effects inherent to effective
Lagrangians, when used to calculate lepton pair production? The
definite example we have in mind is again $\pi + a_1 
\rightarrow e^+ e^-$, evaluated with the help of VMD. We are therefore
concerned by the different possible manifestations of ${\cal L}_{\pi \rho
\, a_1}$. The source of our concern is illustrated as follows. The most
general vertex for $a_1 \rightarrow \pi \rho$ decay can be written as
\begin{eqnarray}
\Gamma^{\mu \nu} \ = \ i g_1 \,g^{\mu \nu} + i g_2\, p^\mu k^\nu + 
i g_3\, p^\nu k^\mu + 
i g_4\, p^\mu p^\nu + i g_5\, k^\mu k^\nu\ ,
\end{eqnarray}
where $g_i (p^2, k^2)$ is a form factor, with 
$p^\mu$ and $k^\nu$ being the the $a_1$ and $\rho$ 
four-momenta. For on-shell $a_1$ decay, three of the above terms are
identically zero, leaving only two.      The form factors are then
chosen such that the experimental width, 
$\Gamma_{a_1 \rightarrow \pi \rho}$, is reproduced. The extra 
experimental constraint is then the
ratio of $D$ to $S$ wave in the final state, which is measured
experimentally \cite{a1ds}. It is probably fair to say that the 
importance of this 
hadronic constraint has not quite penetrated the heavy ion community, 
as revealed by a literature survey. When using the above vertex
for dilepton production, the virtual $\rho$ is pushed off-shell and one
has no direct control over such extrapolations. It turns out  that for
the case at hand, the uncertainty actually resides in the form factors
\cite{gaog97}. We calculate the dilepton rate from the reaction above with:
{\bf (1)} 
\begin{figure}[!hb]
\begin{center}
\includegraphics[angle=90, width=10cm]{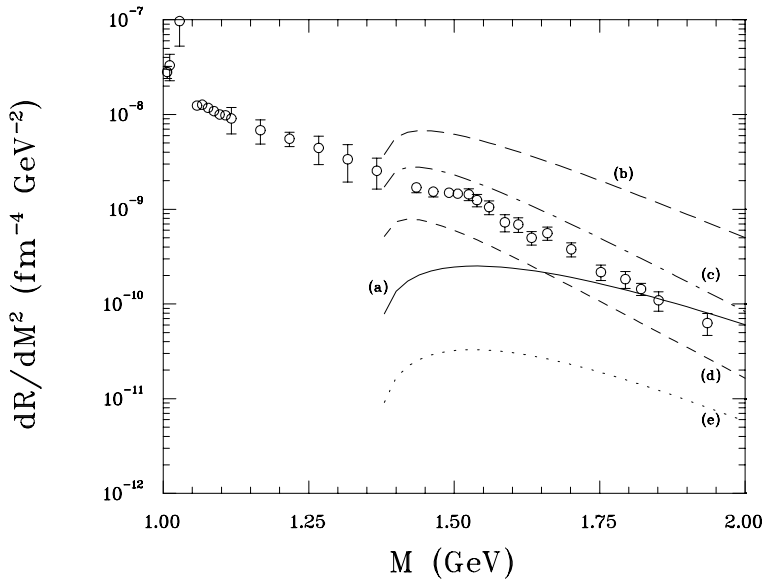}
\end{center}
\caption{\small  Production rates for 
intermediate invariant mass dileptons 
calculated at $T$ = 150 MeV  using  the following interactions:
{\bf (a)} and {\bf (b)} Ko and Rudaz's chiral Lagrangian \cite{ko} (the
difference between {\bf (a)} and {\bf (b)} is explained in the main
text); {\bf(c)} the effective chiral Lagrangian used by Song, 
Ko, and Gale \cite{skg94};{\bf(d)} Lagrangian of Janssen,
Holinde, and  Speth \cite{jhs}; {\bf(e)} Lagrangian used by Xiong, 
Brown, Shuryak \cite{xiong}. Also shown is the rate calculated from
experimentally-fitted spectral functions \cite{huang95} (open circles).} 
\end{figure}
an effective chiral Lagrangian constructed by
Ko and Rudaz based on the $SU(2)_L \times SU(2)_R $
linear $\sigma$ model \cite{ko}; {\bf (2)} an effective chiral 
Lagrangian where the vector mesons are
introduced as massive Yang-Mills fields of the chiral symmetry
\cite{skg94}; {\bf (3)} an effective Lagrangian, previously used by
Xiong, Shuryak and
Brown \cite{xiong} in connection with photon emission rates; {\bf (4)}
an effective Lagrangian developed by Janssen, Holinde, and Speth
\cite{jhs}, to address the issue of form factors in the Bonn potential.
The dilepton rates evaluated with those interactions are displayed in Fig.~2.

The first striking feature of this plot is that the dilepton
rates span three orders of magnitude. Off-shell effects are indeed important,
as also pointed out by a calculation done along similar lines
\cite{kklr96}. For additional help, we can however rely on the 
rates calculated through the  experimentally-extracted spectral 
functions \cite{huang95} used previously. The $\pi + a_1$ reaction
should be a large (if not the largest) contributor to the dilepton signal
in the invariant mass region of interest.  With the available
information we can start our examination of the models. Table 1
summarizes their empirical predictions, along with the actually measured
quantities. 
\begin{table}  
\begin{center}
\begin{tabular}{|c|cc|c|c|c|c|}
\hline
   &\hspace*{1cm} Ko {\em et al.}  &   & Song {\em et al.}    &  Xiong
{\em et al.} & Janssen {\em et al.} &  DATA   \\  
   &  I & II &  & & & \\ 
\hline
$   \Gamma{(a_1\rho \pi)}$ &313.4 & 579.1 & 400  &  400 & 400 &
$\sim 400 $ MeV \\
$   \Gamma{(a_1\pi \gamma) }$& 0.572 & 1.171  & 2.819 &  1.940 &
0.312 & $0.640 \pm 0.246$ MeV \\
$   D/S$ & 0.078 & $-0.168$ & $-0.099$ & 0.185 &  
0.045 &  $-0.09 \pm 0.03$ MeV \\
$\chi^2$ & 11.2 & 9.8 & 39.3 &  37.3 & 11.0 & \\
\hline
\end{tabular} 
\caption{\small The phenomenology of different models.}
\end{center}
\end{table}
Parameter set I for the interaction of Ko {\em et al.} is taken from
Ref.~\cite{kklr96}. Concerning the observables in Table 1, this
parameter set does not actually represent the lowest $\chi^2$: parameter
set II satisfies this criterion. It is interesting to notice that a
modest variation in $\chi^2$ for the fit of the on-shell empirical
observables translates into a large variation of the predicted dilepton
rate. Fig. 2 also contains the rate deduced from the 
experimentally-fitted spectral functions \cite{huang95}. 
Because the process discussed
here is actually included in the spectral function analysis along with
other channels (which are known to interfere constructively), 
the open circles represent a strong upper limit for
the calculations discussed here.  This upper bound is strongly  violated 
by our parameter set II for the Lagrangian of Ref.~\cite{kklr96}. To a
lesser extent this is also the case for the results of calculations with
the interaction of Ref.~\cite{skg94}. In the remaining models in Table
1, the lowest $\chi^2$ is obtained with the Lagrangian of
Ref.~\cite{jhs}. Parameter set I of Ko {\em et al.} has as good a $\chi^2$,
but the slope of the dilepton spectrum departs considerably from that of
the spectral function analysis, with a violation of the upper bound just
below $M$~=~2 GeV. The Lagrangian of Xiong {\em et al.} produces a high
$\chi^2$, with a much flatter slope. We could deem here the interaction of
Janssen {\em et al.} to be the most satisfactory. It can
also be made to saturate the spectral function curve with a proper
choice of form factors, without spoiling its on-shell 
properties. However, the spectral function analysis can be made to be
specific to the $\pi a_1$ channel and the qualitative aspects discussed
here can be set on a quantitative footing; we stick here to our
pedagogical goal and postpone this discussion \cite{gaog97}.

It is clear which caveats will then await us when we embark on the
phenomenological analysis of the experimental data. The purpose of this
exercise was to show that using all of the available empirical
constraints, one  could indeed restore some certainty to an apparently 
ambiguous situation. There are thus reasons to be optimistic about this
type of analyses. The experimentally-extracted spectral function prove
to be helpful, and serve here as baseline calculations. Then, the effective
Lagrangian theory can be used to extrapolate into the realm of
finite-temperature and many-body effects. 

\section*{$\omega-\phi$ mixing at finite temperature}

A last application of effective hadronic Lagrangians will cover here the
mixing of the $\omega$ and $\phi$ vector mesons. 
In vacuum the $\phi$ meson is almost entirely $\bar{s}s$ in its valence
quark content while the $\omega$ meson is almost entirely nonstrange.
There is a small mixing as evidenced by the observed decay mode
$\phi \rightarrow \pi \rho$ and by previous studies using effective hadronic
Lagrangians.  We will study the change in masses and
mixing angles of the $\omega$ and $\phi$ at moderate
temperatures using only conventional ideas.  We consider our study
a simple extrapolation of known physics which may help in deciphering
future experiments.  Specifically, we study the scattering of these
mesons from thermal pions which are the most abundant mesons at
temperatures below 100 MeV or so.  We use effective Lagrangian
techniques to model the relatively soft interactions coupling
pions, $\omega$, and $\phi$ mesons, fitting the parameters to known
physical quantities.  The resulting scattering amplitudes are
used in a virial expansion to compute the vector meson properties
at finite temperature.

A survey of the literature on effective hadronic Lagrangians and
the Review of Particle Physics \cite{revpp} suggests that the
interactions of relevance
involve the 3--point vertices $\phi \rho \pi$, $\omega \rho \pi$,
$\phi b_1 \pi$, and $\omega b_1 \pi$.  The vector self$-$energies
are obtained by computing one loop diagrams at finite temperature
\cite{gsk97}.
The interaction involving the $\rho$ meson is described by the
Wess--Zumino term \cite{WZ}.
\begin{equation}
{\cal L}_{(\omega,\phi) \rho \pi} = g \, \epsilon^{\alpha \beta
\mu \nu} \partial_{\alpha} \mbox{\boldmath $\rho$}_{\beta} \cdot
\mbox{\boldmath $\pi$} \left( \frac{\partial_{\mu} \omega_{\nu}^8
+ \sqrt{2} \, \partial_{\mu} \omega_{\nu}^s}{\sqrt{3}} \right)
\end{equation}
The octet and singlet fields, $\omega_8$ and $\omega_s$,
are expressed in terms of the (vacuum)
physical fields with a mixing angle $\theta_V$.
\begin{eqnarray}
\omega_8 &=& \phi \cos \theta_V + \omega \sin \theta_V \nonumber \\
\omega_s &=& \omega \cos \theta_V - \phi \sin \theta_V
\end{eqnarray}
Ideal mixing is defined such that the physical $\phi$ meson would not
couple to nonstrange hadrons.  This corresponds to $\theta_V =
\theta_{\rm ideal} = \tan^{-1}(1/\sqrt{2}) \approx 35.3^\circ$.  The real
world is not far from that.  Durso \cite{durso}, for example,
fits 39.2$^\circ$, while the Review of Particle Physics quotes 39$^\circ$
based on the Gell--Mann Okubo mass formula.
The coupling constant $g$ is related to the coupling
$g_{VVP}$ used by Gomm, Kaymakcalan, and Schechter \cite{gks} by
$g = - \sqrt{2} \, g_{VVP}$.
Durso
fits $g^2 = 1.62 \pm 0.19 \times 10^{-4}$ MeV$^{-2}$ with a
pseudoscalar mixing angle $\theta_P$ = $- 9.7^\circ$.
We will insure that our parameters reproduce the $\phi
\rightarrow \rho \pi$ decay rate. 
Accepting Durso's value of $g$ we fit $\theta_V = 40.1^\circ$.

The $b_1(1235)$ has a branching ratio of more than 50\% into
$\omega \pi$ and less than 1.5\% into $\phi \pi$.  Thus, the $b_1$
meson is important for the mass shift of the $\omega$ meson at
finite temperature, as noticed by Shuryak \cite{Ed}.  
The interactions are assumed to be SU(3) symmetric
with SU(3) broken only by mass terms.  
The Lagrangian is \cite{gsk97} 
\begin{equation}
{\cal L}_{(\omega,\phi)b_1 \pi} = g_{b_1} \, \mbox{\boldmath $\pi$}
\cdot {\bf b}^{\mu} \left( \frac{ \omega_{\mu}^8
+ \sqrt{2} \, \omega_{\mu}^s}{\sqrt{3}} \right) +
h_{b_1} \, \mbox{\boldmath $\pi$}
\cdot {\bf b}^{\mu\nu} \left( \frac{ \omega_{\mu\nu}^8
+ \sqrt{2} \, \omega_{\mu\nu}^s}{\sqrt{3}} \right) \, .
\end{equation}
The two coupling constants can be inferred from the decay rate
$b_1 \rightarrow \omega \pi$ and from the ratio of the $D$ wave content of
the decay amplitude to its $S$ wave content.
The Review of Particle Physics gives the width 142$\pm$8 MeV and the
measured $D/S$ ratio 0.26$\pm$0.04.  Fitting to central values determines
the most likely values of the coupling constants to be
$g_{b_1} = -9.471$ GeV and $h_{b_1} = 6.642$ GeV$^{-1}$.

The location of the poles and the mixing angle at finite temperature
are obtained by finding the zeros of the inverse propagator in the
(vacuum) physical $\omega$--$\phi$ basis.
\begin{equation}
{\cal D}^{-1}(k_0,{\bf k}) = k_0^2 - {\bf k}^2 - M^2 - \Sigma(k_0,{\bf k})
\end{equation}
$M^2$ is the 2$\times$2 mass matrix at zero temperature.
It is diagonal with components $M^2_{11} = m_{\omega}^2$ and
$M^2_{22} = m_{\phi}^2$.  The self$-$energy has contributions from
the $\rho$ meson and from the $b_1$ meson.  The first was calculated
by Haglin and Gale \cite{HG}.  The second is readily calculated by the usual
finite-temperature rules \cite{Kap}.  We restrict our attention to the 
finite-temperature contribution.  We consider only $\omega$ and $\phi$ mesons
at rest in the many--body system (${\bf k} = 0$) since that is
where temperature will have its maximum impact.  As $|{\bf k}|$
increases, many$-$body effects will decrease, and in the limit
$|{\bf k}| \gg T$ they will disappear altogether.
Finally, the imaginary part of the self$-$energy is small compared to
the real part (inclusive of $M^2$) and so we do not include it in
our present calculation.  We find
\begin{eqnarray}
\Sigma(k_0,|{\bf k}|=0) = \left( \begin{array}{cc}
\cos^2\delta_V & - \left(\cos\delta_V \sin\delta_V \right)/2 \\
- \left(\cos\delta_V \sin\delta_V \right)/2 & \sin^2\delta_V
\end{array} \right)
\left( \Sigma_{\rho} + \Sigma_{b_1} \right) \, ,
\end{eqnarray}
where $\delta_V = \theta_V - \theta_{\rm ideal}$ measures the
deviation from ideal mixing. 
\begin{figure}[!ht]
\begin{center}
\includegraphics[angle=90, width=8cm]{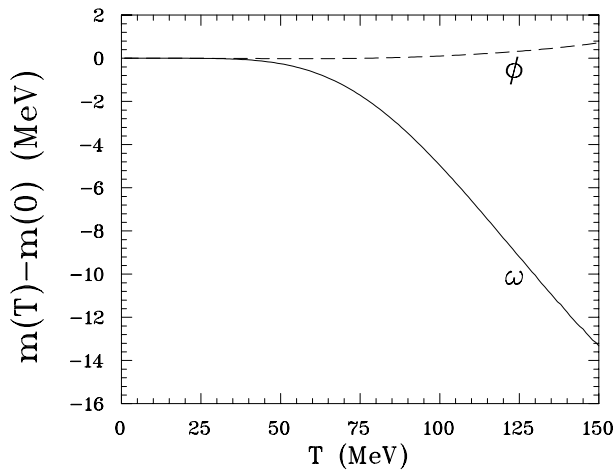}
\end{center}
\caption{\small  Mass shifts of the $\omega$ and $\phi$ 
mesons as functions
of temperature due to scattering from thermal pions.  Multiple pion
scattering and scattering from other thermal mesons should become
important above 100 MeV temperature.} 
\end{figure}
The scalar functions
$\Sigma_{\rho}$ and $\Sigma_{b_1}$ are given in Ref.~\cite{gsk97}.

After diagonalization one finds the mass shifts
as functions of the temperature.  They are displayed in
Fig. 3.  Although the results are plotted
up to a temperature of 150 MeV to see the effect, other scattering
processes will come into play above 100 MeV, as mentioned earlier.
The $\omega$ mass goes down and a shift of roughly 13 MeV is reached at a
temperature of 140 MeV. The $\phi$ mass increases monotonically
to about 0.6 MeV above its vacuum value at a temperature of 140 MeV.
The mixing angles remain uninterestingly small at all temperatures
\cite{gsk97}, but the mass shift of the $\omega$ is potentially
detectable in future experiments.

\section*{Conclusion}

In conclusion, in our brief description of calculations for the low mass
sector we have shown agreement between different theoretical approaches
to dilepton rate calculations. At higher invariant masses, the considerable 
uncertainties associated with off-shell effects have been confirmed.
Nevertheless, we argue that the available phenomenology helps
considerably in narrowing down the possibilities. Finally,  
we compute the finite-temperature change in the masses of the
$\omega$ and $\phi$ mesons and their mixing angle.

\section*{Acknowledgements}

I am happy to acknowledge the collaborators involved with some aspects
of this work.  I am grateful to J. Steele and Z. Huang for
sharing with me their results. 
This work was supported in part by by the Natural Sciences and
Engineering Research Council of Canada, in part by the FCAR fund of the
Qu\'ebec Government, and in part by the US Department of Energy
under grant DE-FG02-87ER40328.

\end{document}